\documentclass[a4paper,aps,prd,onecolumn,preprintnumbers,showpacs,superscriptaddress,nofootinbib]{revtex4}
\usepackage{graphics,graphicx,epsfig}
\usepackage{amsfonts,amsmath,amssymb}

\input{tcilatex}

\begin{document}

\title{Boundary stress tensor and counterterms for weakened AdS$_3$
asymptotic in New Massive Gravity}
\author{Gaston Giribet}
\email{gaston-at-df.uba.ar}
\affiliation{Universidad de Buenos Aires FCEN-UBA and IFIBA-CONICET, Ciudad
Universitaria, Pabell\'on I, 1428, Buenos Aires.}
\author{Mauricio Leston}
\email{mauricio-at-iafe.uba.ar}
\affiliation{Instituto de Astronom\'{\i}a y F\'{\i}sica del Espacio IAFE-CONICET, Ciudad
Universitaria, Pabell\'on IAFE, 1428 C.C. 67 Suc. 28, Buenos Aires.}
\pacs{11.25.Tq, 11.10.Kk}

\begin{abstract}
Resorting to the notion of a stress-tensor induced on the boundary of a
spacetime, we compute the conserved charges associated to exact solutions of
New Massive Gravity that obey weakened versions of AdS$_3$ asymptotic
boundary conditions. The computation requires the introduction of additional
counterterms, which play the r\^ole of regularizing the semiclassical
stress-tensor in the boundary theory. We show that, if treated
appropriately, different ways of prescribing asymptotically AdS$_3$ boundary
conditions yield finite conserved charges for the solutions. The consistency
of the construction manifests itself in that the charges of hairy
asymptotically AdS$_3$ black holes computed by this holography-inspired
method exactly match the values required for the Cardy formula to reproduce
the black hole entropy. We also consider new solutions to the equations of
motion of New Massive Gravity, which happen to fulfill Brown-Henneaux
boundary conditions despite not being Einstein manifolds. These solutions
are shown to yield vanishing boundary stress-tensor. The results obtained in
this paper can be regarded as consistency checks for the prescription
proposed in \cite{Tonni}.
\end{abstract}

\maketitle

\section{Introduction}

Twenty five years have passed since Brown and Henneaux discovered that the
asymptotic dynamics of Einstein gravity in three-dimensional Anti-de Sitter
space (AdS$_{3}$) is generated by the two-dimensional conformal algebra \cite%
{BrownHenneaux}; that is, by two copies of the Virasoro algebra with
non-vanishing central extension. By the middle of 90's, it became clear that
this observation meant much more than an intriguing matching of symmetries.
In 1995, Coussaert, Henneaux, and Van Driel proved that the asymptotic
dynamics of Einstein gravity in AdS$_{3}$ is actually governed by a
two-dimensional conformal field theory (CFT$_2$), at that time identified as
the Liouville field theory \cite{vanDriel}. Later, with the advent of
AdS/CFT correspondence, in 1997, all these observations acquired a natural
framework and were understood from a more general perspective \cite{AdSCFT}.

Over time we learned that the holographic description of the asymptotic AdS$%
_{3}$ dynamics of Einstein gravity in terms of Liouville theory suffers from
some flaws and is not fully satisfactory \cite{CarlipMartinec}; in
particular, in what regards to the statistical description of the
thermodynamical properties of Ba\~{n}ados-Teitelboim-Zanelli black holes 
\cite{BTZ}. However, this did not prevent the experts from addressing the
problem of black hole thermodynamics \cite{Strominger} and other fundamental
problems of quantum gravity \cite{Matt} in terms of the CFT dual
description. In fact, since the beginning, three-dimensional gravity has
proven to be a fruitful testing ground for AdS/CFT correspondence.

More recently, in the last three years, the interest on three-dimensional
gravity in AdS space has been renewed, mainly due to the work of Witten \cite%
{Witten} in which a candidate to be the CFT\ dual of three-dimensional
Einstein's general relativity was presented. The proposal in \cite{Witten}
was that Einstein gravity with negative cosmological constant about AdS$_{3}$
space could be dual to an extremal self-dual two-dimensional conformal field
theory, which exhibits the property of being holomorphically factorizable.
This proposal, in its original form, was subsequently criticized in
different works \cite{Gaberdiel, Gaiotto, Witten2}, in particular in what
concerns to the validity of the construction for large values of the central
charge. Then, it became immediately clear that certain questions on
three-dimensional gravity are far from being clear. One such question is,
for instance, the question about the non-perturbative configurations of
Einstein's gravity, or the question on whether the degenerated and complex
saddle points play any important r\^{o}le at quantum level.

After Witten's paper in 2007, a new proposal for a consistent theory of
quantum gravity in three-dimensional appeared. In \cite{CG}, Li, Song and
Strominger pointed out that, at a very special point of the space of
parameters, the Topologically Massive Gravity \cite{TMG} with negative
cosmological constant seems to lose its local degree of freedom and, at the
same time, the central charge of the left-moving sector of the asymptotic
symmetry algebra vanishes. This observation led the authors of \cite{CG} to
conjecture that, for a specific choice of the coupling constant,
Topologically Massive Gravity is dual to a holomorphic (chiral) conformal
field theory. This proposal is usually referred to as the "Chiral Gravity
Conjecture", and it was extensively discussed in the recent literature \cite%
{DiscussionOnCG}. Subsequently we learned that the realization of the ideas
of \cite{CG} sensibly depends on the way the asymptotic boundary conditions
are prescribed. Actually, this is not surprising; after all, it is well
established that the asymptotic AdS boundary conditions may differ from one
theory to another \cite{HMTZ}, and, besides, a given theory may admit more
than one set of consistent boundary conditions. Therefore, the discussion on
the validity of the proposal in \cite{CG} resulted in a discussion on how to
define what "asymptotically AdS$_{3}$ space" actually means in this context.
This issue was eventually clarified in \cite{Maloney}, where it was pointed
out that two different theories, both defined by the same Lagrangian but
imposing two alternative sets of boundary conditions for each one, seem to
exist. While one of these theories turns out to be dual to a chiral CFT, the
other one, defined by imposing weakened boundary conditions, is believed to
be dual to a Logarithmic CFT \cite{GrumillerJohansson, Maloney, LogTMG}.
This is, indeed, far from being a minor difference, as a Logarithmic CFT is
necessarily non-unitary, giving raise to the question on whether the dual
CFT picture really makes sense if weakened asymptotic conditions are
imposed. Therefore, the lesson we learn from the recent discussions on
three-dimensional chiral gravity was actually instructive: It provides us
with a concise example that shows how dependent on the precise choice of
asymptotic boundary conditions the details of AdS/CFT correspondence can be.

More recently, a new theory of massive gravity in three dimensions was
proposed by Bergshoeff, Hohm, and Townsend \cite{NMG}. This theory, usually
referred to as "New Massive Gravity", also seems to offer a possibility for
formulating a consistent theory of quantum gravity in three dimensions. At
the linearized level, the new theory coincides, after field redefinition,
with the Fierz-Pauli massive model, which turns out to be unitary. Besides,
its action presents other interesting properties \cite{NMG2}-\cite{Tekin}
and exhibits a rich and interesting catalog of solutions \cite{AdSWaves}-%
\cite{Troncoso}. As in the case of Topologically Massive Gravity (TMG), New
Massive Gravity (NMG) also has a point in the space of parameters at which
the central charge of the dual theory vanishes. And the properties of the
dual CFT$_{2}$ also seem to depend on the precise prescription of AdS$_{3}$
boundary conditions. Actually, in NMG there exist more than one way of
relaxing Brown-Henneaux boundary conditions (BH) of three-dimensional
Einstein theory. This is associated to the existence of a massive
parity-invariant graviton modes in the theory. In this paper, we are
interested precisely in studying how relaxing BH asymptotic conditions in
different ways may affect the definition of a regularized stress tensor in
the boundary CFT that would be dual to NMG in AdS$_{3}$ space. More
precisely, we aim to go a step further in the discussion on whether relaxing
BH boundary conditions leads to a well defined CFT in the boundary or not.
Our contribution is to show that, indeed, all the ways of deforming BH
boundary conditions known so far lead to regularizable stress tensor that
can be used to compute conserved charges of asymptotically AdS exact
solutions. Holographic renormalization in NMG\ was recently studied in \cite%
{otro4}.

The paper is organized as follows:\ In Section II, after reviewing the New
theory of Massive Gravity of \cite{NMG}, we discuss the definition of the
boundary stress tensor associated to asymptotically AdS solutions in NMG
recently given in \cite{Tonni}. In Section III, we undertake the computation
of conserved charges of solutions with weakened falling-off in AdS$_{3}$. In
the first subsection of Section III, as a warming up, we discuss the
simplest example of weakened asymptotically AdS boundary conditions: We
consider the logarithmic deformation of the extremal BTZ solution found in 
\cite{GAY, Clement}. This is a solution that emerges at the so-called chiral
point of theory of massive gravity and, despite of that, carries
non-vanishing conserved charges. The mass and angular momentum obtained by
using the boundary stress tensor agree with the results obtained with other
methods in the literature. Then, in the second subsection of Section III, we
consider a different set of relaxed boundary conditions: We analyze the
rotating hairy black hole geometry found in \cite{Troncoso}. For this
geometry, we show that it is actually possible to regularize the boundary
stress tensor by adding local counterterms in the boundary. The need of new
counterterms is due to the soften falling-off that the gravitational field
exhibits, which is the imprint of the gravitational hair. The calculation we
perform yields finite results for both the mass and the angular momentum of
the black hole solution. We explicitly show that the conserved charges
obtained in this way are precisely the ones required for the computation of
the hairy black hole entropy to match with the counting of boundary degrees
of freedom through the Cardy formula. This reinforces the definition of
conserved charges given in \cite{GOTT}. In the third subsection of Section
III, we consider a new solution of NMG that happens to fulfill
Brown-Henneaux boundary conditions but not being an Einstein manifold. This
solution is shown to have vanishing conserved charges when computing with
the same method. Section IV contains our conclusions, which can be
summarized as follows:\ The boundary CFT dual description of NMG\ in
asymptotically AdS$_{3}$ seems to make sense even when relaxed asymptotic
boundary conditions are considered. Alternatively, the consistency of the
results obtained in this paper can be regarded as a non-trivial consistency
check of the holographic prescription proposed in \cite{Tonni}.

\section{Review of New Massive Gravity}

\subsection{Bulk action and equations of motion}

The action of three-dimensional massive gravity is \cite{TMG,NMG} 
\begin{equation}
S=\frac{1}{16\pi G}\int_{\Sigma }d^{3}x\sqrt{-g}\left( R-2\Lambda \right) +%
\frac{1}{32\pi G\mu }\int_{\Sigma }d^{3}x\varepsilon ^{\alpha \beta \gamma
}\Gamma _{\alpha \sigma }^{\rho }(\partial _{\beta }\Gamma _{\gamma \rho
}^{\sigma }+\frac{2}{3}\Gamma _{\beta \eta }^{\sigma }\Gamma _{\gamma \rho
}^{\eta })+\frac{1}{16\pi Gm^{2}}\int_{\Sigma }d^{3}x\sqrt{-g}(R_{\mu \nu
}R^{\mu \nu }-\frac{3}{8}R^{2}).  \label{S}
\end{equation}%
Here we are omitting boundary terms; see (\ref{Sb}) below.

The field equations coming from varying (\ref{S}) with respect to the metric
are 
\begin{equation}
R_{\mu \nu }-\frac{1}{2}Rg_{\mu \nu }+\Lambda g_{\mu \nu }+\frac{1}{2m^{2}}%
K_{\mu \nu }+\frac{1}{\mu }C_{\mu \nu }=0,  \label{fieldeq}
\end{equation}%
where the Cotton tensor $C_{\mu \nu }$ is given by%
\begin{equation}
C_{\mu \nu }=\frac{1}{2}\varepsilon _{\mu }^{\ \text{\ }\alpha \beta }\nabla
_{\alpha }R_{\beta \nu }+\frac{1}{2}\varepsilon _{\nu }^{\ \text{\ }\alpha
\beta }\nabla _{\alpha }R_{\mu \beta },  \label{Cotono}
\end{equation}%
while the tensor $K_{{\mu \nu }}$ is given by 
\begin{equation}
K_{\mu \nu }=2\square {R}_{\mu \nu }-\frac{1}{2}\nabla _{\mu }\nabla _{\nu }{%
R}-\frac{1}{2}\square {R}g_{\mu \nu }+4R_{\mu \alpha \nu \beta }R^{\alpha
\beta }-\frac{3}{2}RR_{\mu \nu }-R_{\alpha \beta }R^{\alpha \beta }g_{\mu
\nu }+\frac{3}{8}R^{2}g_{\mu \nu }.  \label{Kaono}
\end{equation}%
The tensor $K_{\mu \nu }$ obeys the remarkable property $g^{\mu \nu }K_{\mu
\nu }=R_{\mu \nu }R^{\mu \nu }-\frac{3}{8}R^{2}$, i.e. its trace equals the
Lagrangian from which it comes. Cotton tensor, on the other hand, is
traceless, and, in some sense, it can be thought of as the three-dimensional
analogue of the Weyl tensor.

In this paper we are mainly interested in pure New Massive Gravity, namely
the theory defined by setting $\mu =\infty $ in (\ref{S}). Nevertheless,
some of the remarks hold for the general theory. In the case $\mu =\infty $,
an alternative way of writing action (\ref{S}) exists \cite{Tonni}. This
amounts to introduce an second-rank auxiliary field $f_{\mu \nu }$ and
consider the alternative action 
\begin{equation}
S_{\text{A}}=\frac{1}{16\pi G}\int_{\Sigma }d^{3}x\sqrt{-g}\left( R-2\Lambda
+f^{\mu \nu }(R_{\mu \nu }-\frac{1}{2}g_{\mu \nu }R)-\frac{1}{4}m^{2}(f_{\mu
\nu }f^{\mu \nu }-f^{2})\right) ,  \label{Salternativa}
\end{equation}%
where $f=g^{\mu \nu }f_{\mu \nu }$. This permits to turn the problem into
one of second order. It is easy to verify that the equations of motion
derived from (\ref{Salternativa}) coincides with those coming from (\ref{S})
with $\mu =\infty $. In fact, varying (\ref{Salternativa}) with respect to
the non-dynamical field $f_{\mu \nu }$ one finds that the auxiliary field
on-shell is proportional to the Schouten tensor; namely%
\begin{equation}
f_{\mu \nu }=\frac{2}{m^{2}}(R_{\mu \nu }-\frac{1}{4}g_{\mu \nu }R),
\label{schotten}
\end{equation}%
and, then, plugging this into the field equations that come from varying (%
\ref{Salternativa}) with respect to $g_{\mu \nu }$, one recovers equations (%
\ref{fieldeq})-(\ref{Kaono}).

The equations of motion (\ref{fieldeq})-(\ref{Kaono}) admit
three-dimensional Anti-de Sitter space-time (AdS$_{3}$) as exact solutions.
Written in Poincar\'{e} coordinates, the metric of AdS$_{3}$ reads%
\begin{equation}
ds^{2}=-\left( \frac{r^{2}}{l^{2}}+1\right) dt^{2}+\left( \frac{r^{2}}{l^{2}}%
+1\right) ^{-1}dr^{2}+r^{2}d\phi ^{2},  \label{Poincare}
\end{equation}%
where $l$ is the "radius" of the space, which is given in terms of $\Lambda $%
, $\mu $, and $m$. Throughout this paper we will mainly use Poincar\'{e}
coordinates (\ref{Poincare}), which only cover one patch of AdS$_{3}$ space
(this global obstruction is not important for our purpose). In this system
of coordinates, the boundary of the space is located at $r=\infty $ (plus a
point at $r=0$). We will consider the convention such that $x^{0}=t$, $%
x^{1}=\phi $ and $x^{2}=r$, with the Latin indices labeling coordinates $%
i,j=0,1$, and the Greek indices labeling all the coordinates $\mu ,\nu
=0,1,2 $.

In the case of NMG (i.e. $m\neq \infty $ and $\mu =\infty $), we have that
the radius of AdS$_{3}$ may take the values%
\begin{equation}
l_{\pm }^{2}=-\frac{1}{2\Lambda }\left( 1\pm \sqrt{1+\frac{\Lambda }{m^{2}}}%
\right) .  \label{Radius}
\end{equation}%
This shows that NMG has two different maximally symmetric "vacua", each of
them having a different effective cosmological constant $l_{\pm }^{-2}$.

As it happens in the case of Einstein gravity in AdS$_{3}$, the charges
associated to the asymptotic isometries for NMG in AdS$_{3}$ expand two
copies of the Virasoro algebra. The central charge in this case is given by 
\cite{Kraus,Sun}%
\begin{equation}
c=\frac{3l}{2G}\left( 1+\frac{1}{2m^{2}l^{2}}\right) ,  \label{Lac}
\end{equation}%
which reduces to Brown-Henneaux central charge of general relativity \cite%
{BrownHenneaux} in the limit $m\rightarrow \infty $. If $\mu <\infty $, the
theory defined by (\ref{S}) also admits AdS$_{3}$ as a solution. This is
evident once one knows that the Cotton tensor vanishes if and only if the
metric is conformally flat, as AdS$_{3}$ is. Since parity gets broken when $%
\mu <\infty $, it happens that the two copies of Virasoro algebra that
generate the asymptotic diffeomorphisms acquire different central charges.
These are 
\begin{equation}
c_{L}=\frac{3l}{2G}\left( 1-\frac{1}{\mu l}+\frac{1}{2m^{2}l^{2}}\right)
\qquad \text{and}\qquad c_{R}=\frac{3l}{2G}\left( 1+\frac{1}{\mu l}+\frac{1}{%
2m^{2}l^{2}}\right) ,  \label{clcr}
\end{equation}%
which agree with (\ref{Lac}) when $\mu =\infty $. Returning to the case with 
$\mu =\infty $, it is worth mentioning that there exist two special points
in the space of parameters at which the theory exhibits special properties.
One such point is%
\begin{equation}
m^{2}l^{2}=+\frac{1}{2}.  \label{Punto1}
\end{equation}%
At this point, we have $l_{-}^{2}=l_{+}^{2}$, implying that $\Lambda
=-1/(2l^{2})$. This is the point of the moduli space where hairy BTZ-like
black hole solutions are admitted as exact solutions \cite{Troncoso}. Due to
the presence of the gravitational hair, the falling-off of the gravitational
field is weaker than the BH asymptotic exhibited by the BTZ\ solution. In
turn, the question arises as to whether this weakened asymptotic behavior is
also consistent with the existence of a dual CFT\ description in the
boundary. This is precisely the question we want to address in this paper.
We will analyze this in the following Section.

The other point of the space of parameters at which something special
happens is 
\begin{equation}
m^{2}l^{2}=-\frac{1}{2}.  \label{Punto2}
\end{equation}%
At this point, which corresponds to the value $\Lambda =-3/(2l^{2})$, the
central charge (\ref{Lac}) vanishes, resembling what happens in TMG at the
chiral point (i.e. $m=\infty $ with $\mu =1/l$). In fact, one can talk about
a "generalized chiral point" for the theory (\ref{S}), which happens at%
\begin{equation}
\frac{1}{\mu l}-1=\frac{1}{2m^{2}l^{2}},  \label{OOO}
\end{equation}%
i.e. where $c_{L}$ in (\ref{clcr}) vanishes. We will also analyze this point
in the following Section.

\subsection{Boundary action and Brown-York stress tensor}

Now, let us move to discuss the definition of a boundary stress tensor for
NMG in asymptotically AdS$_{3}$ space. This was actually done recently in
Ref. \cite{Tonni}, where the Brown-York type of construction was considered.
The idea, as always one tries to define a holographic stress tensor in this
context, is to start from the Brown-York stress tensor \cite{BY} and then
"push" the whole quantity towards to the boundary. In the process, in
addition to the standard boundary terms, it becomes necessary to add terms
in the action to cancel the divergences that appear at large $r$. These
terms are constructed with intrinsic quantities of the boundary, preserving
the spirit of the holographic correspondence.

Actually, from the perspective of AdS/CFT\ correspondence, one gives new
meaning to the whole idea of constructing a boundary stress tensor in such a
way \cite{BalasubramanianKraus}. In fact, holography provides us with a
physical interpretation for such observable: It can be regarded as the
expectation value $<T_{ij}>$ of the stress tensor of the dual
two-dimensional CFT that is formulated on the boundary. This gives to $%
T_{ij} $ a more concise physical meaning. From this boundary field theory
point of view, the additional terms required to cancel divergences at large $%
r$ are thought of as counterterms, being part of the regularization method.
This holographic stress tensor can then be used, in particular, to compute
conserved charges associated to "localized" solutions in the bulk of AdS$%
_{3} $, or even used to read the value of the central charge of the boundary
CFT$_{2}$. The latter can be accomplished by considering fluctuations of the
boundary metric and computing the expectation value of the trace through the
Weyl anomaly $<T_{i}^{\text{ \ }i}>$ $=\frac{c}{24\pi }\gamma ^{ij}R_{ij}$.
In fact, bulk transformations consistently translate into the anomalous term
in the transformation of $<T_{ij}>$.

The boundary tensor for NMG proposed in \cite{Tonni} was proven to work not
only for asymptotically AdS$_{3}$ metrics, but also for asymptotically
Lifshitz metrics. This is interesting in the context of the recent proposal
for a non-relativistic holography \cite{Kachru}; see also \cite{Ross}.
Nevertheless, in the case of Lifshitz spacetimes it happens that additional
counterterms are needed in order to get a finite result, in contrast to what
happens in the case of AdS$_{3}$ where only a boundary cosmological constant
needs to be added. This suggests that, when trying to define a boundary
stress tensor for configurations of weakened asymptotic in AdS$_{3}$, one
has to be open to the possibility of including additional boundary terms.

Let us review the construction in \cite{Tonni} in more detail: As mentioned, 
\cite{Tonni} follows the standard steps in the construction of a holographic
stress tensor in the context of AdS$_{3}$/CFT$_{2}$. First, boundary terms
are added to the action for the variational principle to be defined in such
a way that it is sufficient to set the variation of the fields to zero in
the boundary. These boundary terms resemble the Gibbons-Hawking term in
Einstein gravity, although the fact the theory under consideration is of
fourth order introduces some differences. Here, we will adopt the
prescription of \cite{Tonni}. Then, once the action is properly defined, one
considers a two-dimensional foliation of the space, defined at constant
values of a preferable radial coordinate $r$. For this purpose it is
convenient to consider the ADM\ decomposition of the metric, in which the
three-dimensional metric on the three-manifold $\Sigma $ reads%
\begin{equation}
ds^{2}=N^{2}dr^{2}+\gamma _{ij}(dx^{i}+N^{i}dr)(dx^{j}+N^{j}dr),  \label{ADM}
\end{equation}%
where $N^{2}$ is the "radial" lapse function, and $\gamma _{ij}$ is the
two-dimensional space-time metric on the constant-$r$ surfaces with
coordinates $x^{i}$ with $i=0,1$.

The next step to construct $T_{ij}$ is to find out the appropriate boundary
terms that come to play the r\^{o}le of the Gibbons-Hawking term of general
relativity. For the case of NMG, this was done in \cite{Tonni} with the help
of the formulation in terms of the auxiliary field $f_{\mu }^{\nu }$. The
boundary action then reads%
\begin{equation}
S_{\text{B}}=\frac{1}{16\pi G}\int_{\partial \Sigma }d^{2}x\sqrt{-\gamma }%
\left( -2K-\widehat{f}^{ij}K_{ij}+\widehat{f}K\right) ,  \label{Sb}
\end{equation}%
where we are using the conventions of \cite{Tonni}; namely, Latin indices
refer to the constant-$r$ surfaces $i,j=0,1$, being $K$ the trace of the
extrinsic curvature, $K=\gamma ^{ij}K_{ij}$. In ADM\ variables (\ref{ADM})
the extrinsic curvature reads $K_{ij}=-\frac{1}{2N}\left( \partial
_{r}\gamma _{ij}-\nabla _{i}N_{j}-\nabla _{j}N_{i}\right) $. The field $%
\widehat{f}_{ij}$ in (\ref{Sb}) is defined as follows 
\begin{equation}
\widehat{f}_{ij}=f_{ij}+2h^{(i}N^{j)}+sN^{i}N^{j}
\end{equation}%
and in the boundary action (\ref{Sb}) we have $\widehat{f}=\gamma ^{ij}%
\widehat{f}_{ij}$.

One can actually verify that the boundary action above leads to a well
defined variational principle; see \cite{Tonni} for details. The boundary
stress tensor is thus defined by varying $S_{\text{A}}+S_{\text{B}}$ with
respect to $\gamma _{ij}$. That is,%
\begin{equation}
T_{ij}=\frac{2}{\sqrt{-\gamma }}\frac{\delta }{\delta \gamma ^{ij}}(S_{\text{%
A}}+S_{\text{B}}).  \label{Tij}
\end{equation}

And it takes the form%
\begin{equation}
16\pi G\text{ }T_{ij}=K_{ij}-\gamma _{ij}K-\frac{1}{2}\widehat{f}%
K_{ij}-\nabla _{(i}\widehat{h}_{j)}\text{ }+\frac{1}{2}\mathcal{D}_{r}%
\widehat{f}_{ij}-K_{k(i}\text{ }\widehat{f}_{j)}^{k}+\frac{1}{2}\widehat{s}%
K_{ij}+\gamma _{ij}(\nabla _{k}\widehat{h}^{k}-\frac{1}{2}\widehat{s}K+\frac{%
1}{2}\widehat{f}K-\frac{1}{2}\mathcal{D}_{r}\widehat{f})  \label{TIJ}
\end{equation}%
where $h^{i}$ and $s$ are given by the components of $f^{\mu \nu }$ that
have at least one radial coordinate, namely $h^{i}=f^{ir}$ (with $i=0,1$)
and $s=f^{rr}$, in terms of which we have also defined $\widehat{h}%
^{ij}=N(h^{i}+sN^{i})$ and $\widehat{s}=N^{2}s$. The covariant radial
derivative $\mathcal{D}_{r}$ in the expression above is defined in such a
way it acts on the fields $\widehat{f}$ and $\widehat{f}^{ij}$ as follows%
\begin{equation}
\mathcal{D}_{r}\widehat{f}_{ij}=\frac{1}{N}\left( \partial _{r}\widehat{f}%
_{ij}-N^{k}\partial _{k}\widehat{f}_{ij}+\widehat{f}_{j}{}^{k}\partial
_{k}N_{i}+\widehat{f}_{i}{}^{k}\partial _{k}N_{j}\right) ,\qquad \mathcal{D}%
_{r}\widehat{f}=\frac{1}{N}\left( \partial _{r}\widehat{f}-N^{k}\partial _{k}%
\widehat{f}\right) .  \label{Dr}
\end{equation}%
Here, in order to make the reading easier, we are using the conventions of
the original reference \cite{Tonni}, to which we refer for the details.

stress tensor (\ref{TIJ}) yields the notion of conserved charges associated
to a given Killing vector $\xi $. Writing the two-dimensional metric as%
\footnote{%
Recall the conventions here:\ $x^{0}=t$, $x^{1}=\phi $, with $i,j=0,1$.} 
\begin{equation*}
\gamma _{ij}\text{ }dx^{i}dx^{j}=-\widehat{N}^{2}dt^{2}+R^{2}(d\phi +%
\widehat{N}_{\phi }dt)^{2},
\end{equation*}%
the charge is defined by \cite{Tonni}%
\begin{equation}
Q[\xi ]=\int d^{2}x\text{ }R\text{ }T_{ij}u^{i}\xi ^{j},
\end{equation}%
where $R$ plays the r\^{o}le of the one-dimensional metric of the constant-$%
t $ lines, and $u^{i}$ is a time-like vector normal to these lines. This
enables to define the mass $\mathcal{M}$ and the angular momentum $\mathcal{J%
}$ of an asymptotically AdS$_{3}$ solution as the conserved charges
associated to the Killing vectors $\xi ^{0}=\partial _{t}$ and $\xi
^{1}=\partial _{\phi }$, respectively. This yields%
\begin{equation}
Q[\partial _{t}]=\mathcal{M}=\int d^{2}x\text{ }R\text{ }T_{i0}u^{i},\qquad
Q[\partial _{\phi }]=\mathcal{J}=\int d^{2}x\text{ }R\text{ }T_{i\phi }u^{i}.
\label{MandJ}
\end{equation}

Now, we are in a position to compute the conserved charges associated to
specific NMG\ solutions. We are specially interested in solutions that,
while being asymptotically AdS$_{3}$ in a sense, do not necessarily obey BH\
boundary conditions. As we anticipated, to accomplish this we will probably
need to add boundary terms to the action (\ref{Sb}) and thus improve the
definition (\ref{Tij})-(\ref{TIJ}) in order to regularize (\ref{MandJ}).
More precisely, we have to supplement the boundary action $S_{\text{B}}$ by
adding an additional piece $S_{\text{C}}$ which would contain quantities
constructed by intrinsic quantities of the boundary. The variation of $S_{%
\text{C}}$ with respect to the boundary metric is what provides the terms
that ultimately cancel the near-boundary divergences. In NMG, because of the
feasibility of formulating the theory in terms of the auxiliary field $%
f_{\mu }^{\nu }$, the set of intrinsic quantities of the boundary among
which we can choose those counterterms gets considerably enhanced with
respect to the case of general relativity. For instance, we have at hand the
following selection of counterterms%
\begin{equation}
S_{\text{C}}=\int d^{2}x\sqrt{-\gamma }\left( \alpha _{0}+\alpha _{1}\hat{f}%
+\alpha _{2}\hat{f}^{2}\right) ,  \label{SC}
\end{equation}%
where the coefficients $\alpha _{i}$ are to be fixed to obtain a finite
result. In turn, the renormalized stress tensor $T_{ij}^{\text{(ren)}}$
turns out to be defined by 
\begin{equation}
T_{ij}^{\text{(ren)}}=T_{ij}+\frac{\delta }{\delta \gamma ^{ij}}S_{\text{C}}.
\label{Tijren}
\end{equation}

For example, for asymptotically AdS$_{3}$ solutions that obey the BH
boundary conditions, it is sufficient to consider a cosmological boundary
term $\alpha _{0}=-\frac{1}{8\pi Gl}(1+\frac{1}{2m^{2}l^{2}})$ with no
additional contributions (i.e. with $\alpha _{1}=\alpha _{2}=0$) to cancel
the divergences \cite{Tonni}. In that case, in turn we have $T_{ij}^{\text{%
(ren)}}=T_{ij}-\frac{1}{8\pi Gl}(1+\frac{1}{2m^{2}l^{2}})\gamma _{ij}$. In
contrast, for asymptotically Lifshitz spaces (with critical exponent $z\neq
1 $) additional terms in (\ref{Tijren}) have to be turned on \cite{Tonni}.
As we will see, even in AdS$_{3}$, if one relax BH\ asymptotic, in general
one has to consider non-vanishing $\alpha _{1}$ and $\alpha _{2}$ to get a
finite result. We will discuss this in the following Section.

\section{Conserved charges and weakened AdS$_{3}$ boundary conditions}

\subsection{A first example:\ Logarithmic deformation of extremal BTZ}

As a warming up, let us begin by considering a simple example of relaxed
boundary conditions. This example was already analyzed in \cite{Tonni}.
Consider the theory (\ref{S}) at the point of the space of parameters such
that $c_{L}=0$. That is, consider the relation (\ref{OOO}). At this point,
there exist exact solutions that are asymptotically AdS$_{3}$ in the sense
of the weakened boundary conditions defined by Grumiller and Johansson in
Ref. \cite{GrumillerJohansson} which, however, do not obey the BH boundary
conditions\footnote{%
More precisely, the solution we consider in this subsection have an
"intermediate" asymptotic; weaker than Brown-Henneaux but still stronger
than Grumiller-Johansson, and consequently consistent with the latter.}. In
fact, it is not hard to verify that equations of motion (\ref{fieldeq})-(\ref%
{Kaono}) admit the following metric as a solution if (\ref{OOO}) holds 
\begin{equation}
ds^{2}=-N^{2}(r)dt^{2}+\frac{dr^{2}}{N^{2}(r)}+r^{2}(N^{\phi }(r)dt-d\phi
)^{2}+N_{k}^{2}(r)(dt-ld\phi )^{2}  \label{unchargedd}
\end{equation}%
where%
\begin{equation}
N^{2}(r)=\frac{r^{2}}{l^{2}}-4GM+\frac{4G^{2}M^{2}l^{2}}{r^{2}},\qquad
\qquad N_{\phi }(r)=\frac{2GMl}{r^{2}},
\end{equation}%
and%
\begin{equation}
N_{k}^{2}(r)=k\log |r^{2}-2GMl^{2}|.  \label{Nkk}
\end{equation}

This metric is a perturbation of the extremal BTZ black hole, which is
recovered setting $k=0$. It can be shown that for $k\neq 0$ the space is
locally equivalent to a $pp$-wave in AdS$_{3}$ space \cite{AdSWaves}. More
properties of the solution were studied in references \cite{GAY, Clement}.

The asymptotic AdS$_{3}$ boundary conditions that the metric (\ref%
{unchargedd})-(\ref{Nkk}) fulfills are given by the following
next-to-leading behavior,%
\begin{eqnarray}
g_{tt} &\simeq &-\frac{r^{2}}{l^{2}}+\mathcal{O}(\log (r)),\qquad
g_{rr}\simeq \frac{l^{2}}{r^{2}}+\mathcal{O}(r^{-4}),\qquad g_{\phi t}\simeq 
\mathcal{O}(\log (r)),  \label{AC} \\
g_{\phi r} &\simeq &\mathcal{O}(1),\qquad g_{\phi \phi }\simeq r^{2}+%
\mathcal{O}(\log (r)),\qquad g_{rt}\simeq \mathcal{O}(1),  \label{AC2}
\end{eqnarray}%
which is evidently weaker than the standard BH boundary conditions. In
particular, the presence of contributions of order $\mathcal{O}(\log (r))$
makes the falling-off of the components $g_{\mu t}$ weaker. Nevertheless,
these boundary conditions are still consistent with the definition of
asymptotic charges that realize the boundary two-dimensional conformal
algebra, and this was studied in \cite{GrumillerJohansson} for the case of
TMG ($m=\infty $, $\mu <\infty $). What we want to emphasize here is that,
despite the weaker falling-off (\ref{AC})-(\ref{AC2}) that the metric
presents, it is still consistent with the definition of a boundary stress
tensor as done in \cite{Tonni}. Moreover, this boundary stress tensor can be
used to calculate the conserved charges associated to the metric (\ref%
{unchargedd})-(\ref{Nkk}) in a very simple way. To see this, let us consider
the case $m^{2}l^{2}=-1/2$ and $\mu =\infty $, for convenience, i.e.
consider $\mu =\infty $ in (\ref{OOO}). Then, since the value of the
boundary cosmological constant required to regularize $T_{ij}$ in NMG\ in AdS%
$_{3}$ happens to be proportional to the quantity $1+\frac{1}{2m^{2}l^{2}}$,
it turns out that at the point (\ref{Punto1}) there is no need to add a
regularizator term in this case, and the computation of the mass and the
angular momentum (\ref{MandJ}) then yields%
\begin{eqnarray}
\mathcal{M} &=&\frac{2k}{G}  \label{19} \\
\mathcal{J} &=&\frac{2k}{G}  \label{20}
\end{eqnarray}%
This agrees with the result obtained by other methods, like the Super
Angular Momentum method considered in \cite{Clement,SAM,WBH}. The
holographic computation of (\ref{19})-(\ref{20}) was already done in \cite%
{Tonni}; here, we have reviewed it as a first working example to emphasize
that, at least in some cases, relaxing the asymptotic may lead to a well
defined $T_{ij}$. However, one could argue that having obtained a finite
result for the charges in this case is not quite surprising since, after
all, (\ref{unchargedd})-(\ref{Nkk}) is a solution that occurs at the chiral
point where special things happen. Therefore, a less simple example would be
to consider solutions that, while exhibiting weakened asymptotic, appear for 
$m^{2}l^{2}\neq -1/2$. Such a solution exists, and we will study it in the
next Subsection.

\subsection{A second example: Hairy Rotating BTZ black hole}

An example of a NMG solution with weakened AdS$_{3}$ asymptotic outside the
chiral point was given in \cite{Troncoso}. This corresponds to a different
deformation of the BTZ geometry that occurs at (\ref{Punto2}), i.e. if $%
m^{2}l^{2}=+1/2$. The metric of the solutions reads

\begin{equation}
ds^{2}=-N(r)F(r)dt^{2}+\frac{dr^{2}}{F(r)}+r^{2}\left( d\phi +N^{\phi
}(r)dt\right) ^{2}\ ,  \label{Hair}
\end{equation}%
where $N(r)$, $N^{\phi }(r)$ and $F(r)$ are functions of the radial
coordinate $r$, given by%
\begin{align}
N(r)& =\left[ 1+\frac{bl^{2}}{4H(r)}\left( 1-\eta \right) \right] ^{2}\
,\qquad N^{\phi }(r)=-\frac{a}{2r^{2}}\left( 4GM-bH(r)\right) \ ,  \notag \\
F(r)& =\frac{H^{2}(r)}{r^{2}}\left[ \frac{H^{2}(r)}{l^{2}}+\frac{b}{2}\left(
1+\eta \right) H(r)+\frac{b^{2}l^{2}}{16}\left( 1-\eta \right) ^{2}-4GM\
\eta \right] \ ,  \label{hairy}
\end{align}%
and 
\begin{equation}
H(r)=\left[ r^{2}-2GMl^{2}\left( 1-\eta \right) -\frac{b^{2}l^{4}}{16}\left(
1-\eta \right) ^{2}\right] ^{\frac{1}{2}}\ .  \label{H}
\end{equation}%
Here\footnote{%
The parameter $\eta $ was introduced here, and it relates to the notation
used in \cite{GOTT} by $\eta =\Xi ^{1/2}$.} $\eta =\sqrt{1-a^{2}/l^{2}}$,
and $a=J/M$ is the rotation parameter. Actually, for certain range of the
parameters $M,$ $J,$ and $b$ this solution represents a hairy rotating black
hole. The parameter $b$ represents the "gravitational hair" of the solution,
and for the solution to be a black hole the bounds $M>-\frac{b^{2}l^{2}}{16G}
$ and $-Ml\leq J\leq Ml$ need to be satisfied.

For computational purposes it may result convenient to redefine the radial
coordinate as $r\rightarrow \widehat{r}=H(r)$. This does not change the
asymptotic since near the boundary $r\simeq \widehat{r}+\mathcal{O}(1/r)$.
It is not hard to verify that the solution is asymptotically AdS$_{3}$ in a
weak sense. More precisely, it obeys the following boundary conditions

\begin{eqnarray}
g_{tt} &\simeq &-\frac{r^{2}}{l^{2}}+\mathcal{O}(r),\qquad g_{rr}\simeq 
\frac{l^{2}}{r^{2}}+\mathcal{O}(r^{-3}),\qquad g_{\phi t}\simeq \mathcal{O}%
(r),  \label{UO} \\
g_{\phi r} &\simeq &\mathcal{O}(1),\qquad g_{\phi \phi }\simeq r^{2}+%
\mathcal{O}(1),\qquad g_{rt}\simeq \mathcal{O}(1),  \label{OU}
\end{eqnarray}

Notice that these asymptotic conditions are not the ones in (\ref{AC})-(\ref%
{AC2}), nor agree with the BH asymptotic conditions. This is the key point
here:\ We will prove that this (new version of) relaxed asymptotic also
yields a well defined stress tensor in the boundary. But, first, let us
discuss a little more about the geometry (\ref{Hair})-(\ref{H}). The
properties of the hairy black hole solution were studied in \cite{Troncoso}
and \cite{GOTT}. In particular, its thermodynamics was studied: The solution
has Hawking temperature%
\begin{equation}
T_{\text{H}}=\frac{\eta }{\pi l}\sqrt{2G\left( M+\frac{b^{2}l^{2}}{16G}%
\right) \left( 1+\eta \right) ^{-1}},  \label{Temperature}
\end{equation}%
and Bekenstein-Hawking entropy%
\begin{equation}
S_{\text{BH}}=\pi l\sqrt{\frac{2}{G}\left( M+\frac{b^{2}l^{2}}{16G}\right)
\left( 1+\eta \right) }.  \label{Entropy}
\end{equation}%
These quantities fulfill the relation

\begin{equation}
T_{\text{H}}\text{ }dS_{\text{BH}}=\eta \text{ }dM+\frac{bl^{2}}{8G}\eta 
\text{ }db-\frac{1}{a}\left( 1-\eta \right) \left( M+\frac{b^{2}l^{2}}{16G}%
\right) \text{ }da\ ,  \label{calentate}
\end{equation}%
where%
\begin{equation}
\Omega _{+}=\frac{1}{a}\left( \eta -1\right) \   \label{Omega+}
\end{equation}%
is the angular velocity of the horizon. We will see below how the correct
definition of the black hole mass and angular momentum yields both a
statistical explanation for (\ref{Entropy}) and a physical meaning for (\ref%
{calentate}).

Then, in principle we are ready to compute the conserved charges associated
to this black hole solution resorting to the definition of (\ref{Tijren}).
However, when evaluating the integrals in (\ref{MandJ}) one rapidly notices
that, because of the weaker falling-off (\ref{UO})-(\ref{OU}), the
divergences are severer than in the case of BH boundary conditions. This
demands to consider more counterterms than a mere boundary cosmological
constant term. Besides, it entails an extra difficulty since one also has to
analyze the (non)ambiguity of the definition of the charges by choosing
different prescriptions to regularize. For instance, if one includes the
three terms of (\ref{SC}) and first tries to calculate the mass of the black
hole solution (\ref{Hair})-(\ref{H}) for the particular case $a=0$, then one
finds only two conditions for the three couplings $\alpha _{i}$, namely $%
\alpha _{0}=16\alpha _{2}$ and $\alpha _{1}=1/l+8\alpha _{2}$, and this
introduces an undesired ambiguity in the value of the mass. However, one can
show that such ambiguity disappears when the rotating solution (i.e. $a\neq 0
$)\ is considered. In that case, demanding the finiteness of the result
completely fixes the three coefficients to be $\alpha _{0}=\alpha _{2}=0$
and $\alpha _{1}=1/l$, and the computation of (\ref{MandJ}) using the
improved stress tensor (\ref{Tijren}) yields%
\begin{equation}
\mathcal{M}=M+\frac{b^{2}l^{2}}{16G}\ .  \label{LaM}
\end{equation}%
\begin{equation}
\mathcal{J}=J-\frac{ab^{2}l^{2}}{16G}.  \label{TheJ}
\end{equation}

It is worth noticing that these values for the charges, considered together
with (\ref{calentate}), verify the first principle of black hole
thermodynamics, which reads%
\begin{equation}
d\mathcal{M}=T_{\text{H}}\text{ }dS_{\text{BH}}-\Omega _{+}\text{ }d\mathcal{%
J}\ .
\end{equation}

Furthermore, we can verify that the result obtained for the conserved
charges in (\ref{LaM})-(\ref{TheJ}) is exactly the one required for the dual
CFT$_{2}$ to account for the black hole entropy. Let us summarize this story
here: The first precise observation about the statistical counting of the
three-dimensional\ black hole degrees of freedom in terms of its boundary
dual theory was made by Strominger in Ref. \cite{Strominger} for the case of
Einstein gravity. Strominger noticed that, in virtue of the results of \cite%
{BrownHenneaux}, the formula for the density of states of the boundary CFT$%
_{2}$ exactly reproduces the Bekenstein-Hawking entropy of the BTZ\ black
hole. The computation uses the fact that, for any CFT$_{2}$ that satisfies
some physically sensible requirements\footnote{%
Cardy formula strongly relies on modular invariance at one loop, and on
certain assumptions on the gap in the spectrum that permit to use a saddle
point approximation. See \cite{CarlipMartinec} for a discussion on Cardy
formula in this context.}, the density of states $\rho (h,\overline{h})$ of
a given conformal dimension ($h,\overline{h}$) asymptotically grows
following a very simple expression, called the Cardy formula \cite{Cardy}.
More precisely,%
\begin{equation}
\rho (h,\overline{h})\simeq e^{2\pi \sqrt{\frac{ch}{6}}}e^{2\pi \sqrt{\frac{c%
\overline{h}}{6}}},  \label{Log}
\end{equation}%
where $c$ is the central charge of the theory. This, in turn, yields a very
simple expression for the entropy in the microcanonical ensemble; namely%
\begin{equation}
S_{\text{CFT}}=2\pi \sqrt{\frac{ch}{6}}+2\pi \sqrt{\frac{c\overline{h}}{6}}.
\label{Cardy2}
\end{equation}

The quantities $\mathcal{E}=h+\overline{h}$ and $\mathcal{J}=h-\overline{h}$
are typically associated to the energy (in units of $1/l$) and the spin of
the fields in the CFT$_{2}$, respectively. In the dual description these
quantities turn out to be in correspondence with the mass and angular
momentum of the asymptotically AdS$_{3}$ black holes states.

The observation made in \cite{Strominger} was that, if one considers the
Brown-Henneaux central charge $c=\frac{3l}{2G}$ for general relativity, and
identify the mass and the angular momentum of an asymptotically AdS$_{3}$
solution appropriately, then Cardy formula (\ref{Cardy2}) happens to match
the entropy of the BTZ\ black hole. This is a remarkable observation, which
merely relies on general aspects of the asymptotic symmetry of Einstein
gravity in AdS$_{3}$. What we want to point out here is that, if we relate
the mass (\ref{LaM}) and the angular momentum (\ref{TheJ}) to the conformal
dimension as%
\begin{equation}
l\mathcal{M}=h+\overline{h},\qquad \mathcal{J}=h-\overline{h},
\label{LasCargas}
\end{equation}%
then (\ref{Cardy2}) exactly reproduces the entropy of the hairy rotating
black hole too; see also \cite{GOTT} for analogous computation. This is
quite remarkable since, in contrast to BTZ\ black hole, the metric of the
hairy black hole solution (\ref{Hair}) does not satisfy the Brown-Henneaux
boundary condition, but a relaxed version of them. Actually, considering
that for (\ref{Punto1}) the central charge takes the value\footnote{%
Notice this is twice the Brown-Henneaux central charge for general
relativity.} 
\begin{equation*}
c=\frac{3l}{G},
\end{equation*}%
\ and putting together (\ref{LaM}), (\ref{TheJ}), (\ref{Cardy2}), and (\ref%
{LasCargas}), one finally recovers the entropy (\ref{Entropy}). Namely,%
\begin{equation}
S_{\text{BH}}=S_{\text{CFT}} .  \label{Agreement}
\end{equation}

It is worth emphasizing that having obtained this matching is not trivial:
If the second term in (\ref{LaM}) (and/or in (\ref{TheJ})) were not enter in
the definition of the black hole mass (and angular momentum), then Cardy
formula would have not reproduced the Bekenstein-Hawking entropy, cf. \cite%
{Troncoso}. This is why having obtained such a dependence on $b$ in the mass
formula is pleasant. The argument in \cite{GOTT} to include such a $b$%
-dependent term was that the absence of a chemical potential associated to
the hair parameter $b$ makes possible to absorb its variation by redefining
the conserved charges in the first principle of black hole thermodynamics.
This naturally leads to consider the solution with $M+\frac{b^{2}l^{2}}{16G}%
=0$ as the "ground state". Here, we have obtained this result in a
completely independent way, confirming that the conserved charges (\ref{LaM}%
) and (\ref{TheJ}) are the correct result, and thus completing the argument
of \cite{GOTT}.

\subsection{A third example:\ Non-Einstein deformation of BTZ geometry with
Brown-Henneaux asymptotic}

So far, we have considered two different deformations of the BTZ\ geometry,
each of them representing different ways of relaxing Brown-Henneaux boundary
conditions. We have seen how, in both cases, a boundary stress tensor can be
actually defined and used to compute finite conserved charges. While
solution (\ref{unchargedd})-(\ref{Nkk})\ satisfies boundary conditions (\ref%
{AC})-(\ref{AC2}) and did require no counterterms for its charges (\ref{19}%
)-(\ref{20}) to be computed, solution (\ref{Hair})-(\ref{H}) satisfies the
asymptotic (\ref{UO})-(\ref{OU})\ and has conserved charges (\ref{LaM})-(\ref%
{TheJ}) which were computed by regularizing the stress tensor. Now, let us
consider a third way of perturbing the BTZ\ geometry; one preserving BH
asymptotic but not being a solution of general relativity. Consider again
the point of the parameter space (\ref{Punto1}), at which the central charge
(\ref{Lac}) vanishes. And consider the following perturbation of the
zero-mass BTZ\ solution (here we set $l=1$) 
\begin{equation}
ds^{2}=-r^{2}dt^{2}+\frac{dr^{2}}{r^{2}}+r^{2}d\phi ^{2}+N_{\gamma
}^{2}(dt+d\phi )^{2}  \label{A}
\end{equation}%
with 
\begin{equation}
N_{\gamma }^{2}=\left( \alpha +\beta t+\frac{\gamma }{r^{4}}\right) .
\label{Ng}
\end{equation}%
It can be verified that this \textit{ansatz} solves the field equations (\ref%
{fieldeq})-(\ref{Kaono}) for $m^{2}l^{2}=-1/2$, $\mu =\infty $, and if%
\begin{equation}
\gamma =-\frac{\beta ^{2}}{72}.  \label{ammabeta}
\end{equation}%
The metric exhibits the following asymptotic conditions%
\begin{eqnarray}
g_{tt} &=&-r^{2}+\alpha +\beta t+\frac{\gamma }{r^{4}},\qquad g_{rr}=\frac{1%
}{r^{2}}  \label{conditionsA} \\
g_{\phi \phi } &=&r^{2}+\alpha +\beta t+\frac{\gamma }{r^{4}},\qquad g_{\phi
t}=\alpha +\beta t+\frac{\gamma }{r^{4}},  \label{conditionsNg}
\end{eqnarray}%
where $\alpha ,\beta ,\gamma $ are real numbers. In turn, this represents a
NMG solution that fulfills Brown-Henneaux boundary conditions. What is
remarkable is that the metric (\ref{A})-(\ref{Ng}) is an asymptotically AdS$%
_{3}$ solution of NMG\ in the sense of BH but it is not an Einstein manifold%
\footnote{%
This solution was first found by S. de Buyl, G. Comp\`{e}re, and S.
Detournay for the case of TMG\ at the chiral point \cite{belgas}. We found
the NMG\ analogue inspired in their unpublished work, where the properties
of the geometry and its relevance for the chiral gravity conjecture are
analyzed.}. Interestingly enough, this localized solution yields an
expression for the (unrenormalized) stress tensor (\ref{TIJ}) that vanishes
identically. Analogous result is obtained (\ref{A})-(\ref{Ng}). A similar
solution was found for the case of chiral gravity in \cite{belgas}; we refer
to that work for a discussion on the properties of (\ref{A})-(\ref{Ng}) and
its physical relevance. Let us say here that the fact $T_{ij}=0$ when
evaluated on (\ref{A})-(\ref{Ng}) could be relevant for the discussion on
which are the sectors to be taken into account for computing a partition
function in massive gravity on AdS$_{3}$ at the point $c=0$. This is because
(\ref{conditionsA})-(\ref{conditionsNg}) are consistent with the BH
asymptotic \cite{BrownHenneaux} even when metric (\ref{A})-(\ref{Ng}) is not
locally AdS$_{3}$, and consequently it raises the question whether this
solution should be included or not as a saddle point of the AdS$_{3}$
sector. The fact such a solution of massive gravity carry vanishing charges
is interesting.

\section{Conclusions}

In this paper we have considered different solutions of NMG in
asymptotically AdS$_{3}$ spaces, each of them incarnating a different
notions of what "asymptotically AdS$_{3}$" means. All the solutions
discussed here represent different ways of deforming the BTZ geometry. We
considered both solutions satisfying Brown-Henneaux boundary conditions and
solutions with relaxed asymptotic, having computed the conserved charges
associated to all of them. In particular, we studied the logarithmic
deformation of the BTZ geometry \cite{GAY, Clement} that appears at the
chiral point, which obeys the asymptotic conditions proposed by Grumiller
and Johansson in \cite{GrumillerJohansson}. We also considered the hairy
rotating black hole solution of \cite{Troncoso}, for which we reconsidered
its thermodynamics in light of the holographic computation of the conserved
charges. We also studied a new solution to NMG, which obeys BH asymptotic
not being a solution of Einstein gravity. For all the asymptotically AdS$%
_{3} $ solutions studied in this paper it was possible to define a boundary
stress tensor; even for those that do not exhibit BH asymptotic. It is
likely that NMG formulated in asymptotically AdS$_{3}$ space, considering
weakened version of BH boundary conditions, is dual to a two-dimensional
conformal field theory. In fact, evidence suggesting that AdS/CFT
correspondence resists such a relaxation of the asymptotic conditions
exists. In the literature we find the following two suggestive observations:

\begin{itemize}
\item The group of asymptotic symmetry defined with weaker version of
Brown-Henneaux boundary conditions is also generated by two copies of
Virasoro algebra with a central extension. The central charge turns out to
be exactly the same as if Brown-Henneaux conditions had been imposed \cite%
{Sun,Troncoso}; although the other properties of the CFT$_{2}$ hardly remain
unaltered by the change in the boundary conditions.

\item The structure exhibited by two- and three-point functions with relaxed
falling-off at $c=0$ seem to be consistent with the functional form of a
(presumably Logarithmic) conformal field theory \cite{LogNMG}.
\end{itemize}

In this paper, we added the following piece of evidence:

\begin{itemize}
\item It is possible to define a boundary stress tensor for NMG\ in AdS$_{3}$
even if relaxed asymptotic is considered. In contrast to the case of
Brown-Henneaux boundary conditions, additional counterterms are needed to
regularize the divergences in the stress tensor. These divergences are
induced by the soften falling-off of the bulk gravitational field. The
conserved charges computed with the regularized stress tensor exactly match
the values required for the Cardy formula in the boundary CFT$_{2}$ to
reproduce the Bekenstein-Hawking entropy of the hairy black hole in AdS$_{3}$%
.
\end{itemize}

All this seems to suggest that a dual CFT$_{2}$ description still exists if
weakened asymptotic is prescribed. As it happens in TMG\ at the chiral
point, the cost of relaxing AdS$_{3}$ boundary conditions in NMG\ may be
that of losing unitarity in the dual CFT$_{2}$. This raises the question on
whether it makes sense to consider such a way of formulating AdS/CFT.
However, from a less conservative point of view, one can argue that AdS/CFT\
correspondence could make sense even for non-unitary CFTs. After all, in
condensed matter applications non-unitary CFTs play an important r\^{o}le,
and having a gravity dual for Logarithmic conformal field theory could be
very interesting in this context.

Let us emphasize that having found appropriate counterterms to regularize $%
<T_{ij}>$ in the case of weakened AdS$_{3}$ asymptotic is non-trivial, as a
priori there is no guarantee to achieve so. A good example is given by
trying to proceed in the same way for the case of the asymptotically
Warped-AdS$_{3}$ spaces in NMG. In fact, there is no obvious manner of
defining a regularized boundary stress tensor in that case. In TMG, for
which the Brown-York construction was studied in \cite{BalasubramanianKraus}%
, the situation is actually similar. However, the fact of not being able to
fully regularize all the components of $T_{ij}$ should not prevent us from
employing the holographic-inspired method to compute at least some conserved
quantities of asymptotically Warped-AdS$_{3}$ solutions. To be more precise,
let us illustrate this by analyzing the case of asymptotically Warped-AdS$%
_{3}$ black holes in TMG \cite{WBH}. The metric of these black holes is \cite%
{WAdS3,GeoffreyStephane,Dio} 
\begin{eqnarray}
\frac{ds^{2}}{l^{2}} &=&dt^{2}+\frac{dr^{2}}{(\nu ^{2}+3)(r-r_{+})(r-r_{-})}%
+\left( 2\nu r-\sqrt{r_{+}r_{-}(\nu ^{2}+3)}\right) dtd\theta +  \notag
\label{CWADS} \\
&&\frac{r}{4}\left( 3(\nu ^{2}-1)r+(\nu ^{2}+3)(r_{+}+r_{-})-4\nu \sqrt{%
r_{+}r_{-}(\nu ^{2}+3)}\right) d\theta ^{2}
\end{eqnarray}%
where $r_{\pm }$ are the location of the horizons of the rotating solution;
here we use the conventions of Ref. \cite{WAdS3}, defining $\nu =\mu l/3$
and considering now $m=\infty $. It is possible to show that, by adding a
boundary cosmological counterterm\footnote{%
Notice that the value of the boundary cosmological constant matches the
value $1/l$ for the case $\nu =1$, where the Warped-AdS$_{3}$ space
coincides with AdS$_{3}$ space. The introduction of a boundary cosmological
constant to regularize the boundary stress-tensor for asymptotically
Warped-AdS$_{3}$ spaces was also considered by Daniel Grumiller and Niklas
Johansson. The authors thank Alan Garbarz for suggesting this idea to them.}%
\begin{equation}
S_{\text{C}}=-\frac{\sqrt{\nu ^{2}+3}}{16\pi Gl}\int d^{2}x\sqrt{-\gamma },
\label{SCC}
\end{equation}%
one obtains a finite result for the mass of the Warped-AdS$_{3}$ black hole.
Remarkably, the finite result found for the mass happens to be the correct
value, which in the conventions of \cite{WAdS3} reads%
\begin{equation}
\mathcal{M}=\frac{(\nu ^{3}+3)}{48Gl}\left( r_{+}+r_{-}-\frac{1}{\nu }\sqrt{%
r_{+}r_{-}(\nu ^{2}+3)}\right) .
\end{equation}

This formula is cumbersome enough for not to doubt about that the
calculation of this particular component does make sense. However, the
introduction of (\ref{SCC}) does not suffice to regularize the angular
momentum. Our failed attempts to find a natural way of defining a fully
regularized boundary stress tensor for asymptotically Warped-AdS$_{3}$
spaces, together with the difficulties encountered in trying to derive one
of the two central charges of the asymptotic algebra \cite{GeoffreyStephane}%
, suggest that probably there is no dual conformal field theory description
for asymptotically Warped-AdS$_{3}$ spaces, at least not in an orthodox way.
This shows that having been able to regularize (\ref{TIJ}) for weakened AdS$%
_{3}$ asymptotic resorting to (\ref{SC}) is actually non-trivial.

Furthermore, we would like to argue that having obtained a result for the
hairy black hole mass that turns out to be consistent with the Cardy formula
can also be considered as a consistency check of the holographic
prescription proposed in \cite{Tonni}. More precisely, let us be reminded of
the fact that in the definition of the regularized stress tensor giving in 
\cite{Tonni} there existed an ambiguity in choosing which is the auxiliary
field that has to be considered as "fundamental" when varying with respect
to the boundary metric. Even though this ambiguity obviously does not affect
the theory in the bulk, it does affect the definition of $T_{ij}$ and,
consequently, the correct values of the conserved charges. In turn, the fact
that we found the agreement (\ref{Agreement}) convinces us that the result (%
\ref{LaM})\ is the correct value and that the prescription of \cite{Tonni}
seems to be physically sensible even when weakened boundary conditions are
considered.

Another aspect of the computation of the hairy black hole entropy we find
interesting is the following: Unlike what happens in the non-rotating hairy
solution (i.e. $a=0$ in (\ref{Hair})) for which the choice of counterterms
coefficients $\alpha _i$ presents some ambiguity that ultimately translates
into an ambiguity in the value of the mass, in the case of the rotating
solution ($a \neq 0$) such ambiguity disappears as the three coefficients $%
\alpha _i$ get completely fixed yielding specific values for both ${\mathcal{%
M}}$ and ${\mathcal{J}}$. That is, if one only knew the non-rotating
solution of the theory, then the holographic computation of the black hole
quantities would be ambiguous, and it is only when the angular momentum is
turned-on that the conserved charges get fully determined. This is like
saying that the computation only works well once one has identified the
complete family of solutions to which the member one is interested in
belongs. Likely, a rotating version of the asymptotically Lifshitz$_{z=3}$
black hole of NMG \cite{Lifshitz} also exits, and finding this solution
would lead to completely fix the ambiguity in the choice of counterterms for
calculating its mass.

Going back to the case of AdS$_{3}$: The main result of this paper was
showing that the definition of a boundary stress tensor proposed in \cite%
{Tonni} can be extended to the case of weakened AdS$_{3}$ asymptotic. Our
computation turns out to be in accordance with AdS/CFT, whose validity seems
to be robust under the relaxation of the falling-off of the gravitational
field. Nevertheless, further evidence is needed to confirm the
interpretation of (\ref{TIJ}) as being the stress tensor of a dual
two-dimensional conformal field theory. For instance, the question remains
whether the proposed expression for the boundary $<T_{ij}>$ can be used to
calculate other quantities of the boundary theory, like correlation
functions. This could be particularly important for the theory at the
special point $c=0$, which has been conjectured to be a Logarithmic CFT$_{2}$%
, and the two-point function of the stress tensor would then exhibit a
special form. Studying this and other aspects of the dual conformal model in
the case of weakened asymptotic is matter of further study.

\begin{equation*}
\end{equation*}

This work was supported by UBA, CONICET and ANPCyT. The authors thank the
following colleagues for useful discussions on related subjects: Daniel
Grumiller, Niklas Johansson, Olivera Miskovi\'{c}, Rodrigo Olea, Massimo
Porrati, and Ricardo Troncoso. The authors are grateful to Sophie de Buyl,
Geoffrey Comp\`{e}re, St\'ephane Detournay, Alan Garbarz, Olaf Hohm, and
Erik Tonni for collaboration in this subject.

\bigskip

\end{document}